\documentclass[aps,pra,10pt,a4paper,reprint,nofootinbib,superscriptaddress]{revtex4-2}
\usepackage{xcolor}
\usepackage[utf8]{inputenc}
\usepackage{amsfonts}
\usepackage{amssymb}
\usepackage{amsmath}
\usepackage{amsthm}
\usepackage{graphics}
\usepackage{graphicx}
\usepackage{braket}
\usepackage[normalem]{ulem}
\usepackage[colorlinks=true,linkcolor=blue,urlcolor=blue,citecolor=blue]{hyperref}
\usepackage{times,txfonts}
\newtheorem{theorem}{Theorem}

\theoremstyle{definition}
\newtheorem{example}{Example}

\def\Tr{\text{Tr}}

\begin{document}


\title{Enhanced quantum state discrimination under general measurements\\with entanglement and nonorthogonality restrictions}

\author{Swati Choudhary}
\affiliation{Harish-Chandra Research Institute,  A CI of Homi Bhabha National Institute, Chhatnag Road, Jhunsi, Prayagraj  211 019, India}
\affiliation{Center for Quantum Science and Technology (CQST) and  
Center for Computational Natural Sciences and Bioinformatics (CCNSB),
International Institute of Information Technology Hyderabad, Prof. CR Rao Road, Gachibowli, Hyderabad 500 032, Telangana, India}

\author{Aparajita Bhattacharyya}
\affiliation{Harish-Chandra Research Institute,  A CI of Homi Bhabha National Institute, Chhatnag Road, Jhunsi, Prayagraj  211 019, India}

\author{Ujjwal Sen}
\affiliation{Harish-Chandra Research Institute,  A CI of Homi Bhabha National Institute, Chhatnag Road, Jhunsi, Prayagraj  211 019, India}

\begin{abstract}
The minimum error probability for distinguishing between two quantum states is bounded by the Helstrom limit, derived under the assumption that measurement strategies are restricted to positive operator-valued measurements. We explore scenarios in which the error probability for discriminating two quantum states can be reduced below the Helstrom bound under some constrained access of resources, indicating the use of measurement operations that go beyond the standard positive operator-valued measurements  framework. We refer to such measurements as non-positive operator-valued measurements. While existing literature often associates these measurements with initial entanglement between the system and an auxiliary, followed by joint projective measurement and discarding the auxiliary, we demonstrate that initial entanglement between system and auxiliary is not  necessary for the emergence of such measurements in the context of state discrimination. Interestingly, even initial product states can give rise to effective non-positive measurements  on the subsystem, and achieve sub-Helstrom discrimination error when discriminating quantum states of the subsystem.
\end{abstract}

\maketitle

\section{Introduction}\label{sec:Introduction}
Quantum state discrimination (QSD) is the process of identifying an unknown quantum state chosen according to a known probability distribution from a finite set of possible states~\cite{C2010,JUM2004,JL2015,SS2009,J2010}. This task is achieved through appropriate quantum measurements on the given state. QSD plays a crucial role in a wide range of quantum information protocols and technologies, including quantum channel discrimination~\cite{M2001,M2005,GGP2008,A2014,AM2016,MM2018,M2020}, random access codes~\cite{AATU1999,A1999,ADLM2008,GNJ2023,PA2023}, quantum dense coding~\cite{CS1992,T2001,XDGF2002,MV2003,DGMCAU2004,SJB2005,MM2012,CAAU2019}, and quantum metrology~\cite{M2009,BRL2011,VSL2011,GI2014,LAMR2018}. Consequently, it is of significant importance not only for practical applications but also for foundational studies in quantum theory.

The task of discriminating quantum states is inherently constrained by the fundamental principles of quantum mechanics. For instance, the no-cloning theorem prohibits the perfect replication of an arbitrary unknown quantum state~\cite{WW1982}. As a result, perfect discrimination between non-orthogonal states using quantum operations alone is fundamentally impossible. Even in scenarios where the candidate states are mutually orthogonal, perfect discrimination may not be achievable when the state is distributed among multiple parties restricted to local operations and classical communication~\cite{CDC++1999}. QSD strategies can generally be categorized into two main approaches: Minimum-Error State Discrimination~\cite{UJ2002,MJ2002,CL2003,UJ2004,D2008,DL2010,AGG2010,YRDSK2010,MMOAL2017,E2022,DJ2022} and Unambiguous State Discrimination ~\cite{AS1998,MMN2000,A2001,YMJ2001,YJM2002,S2002,Y2003,MAUK2003,YRM2004,UJ2005,JM2005,BJ2006,MMNA2008,MHD2010,MHD22010,S2014,SL++2021}. In Minimum-Error State Discrimination, a guess must always be made regarding the identity of the state, resulting in an inherent probability of error. The objective is to minimize this error. In contrast, Unambiguous State Discrimination allows for inconclusive outcomes, wherein no guess is made in some instances; however, whenever a conclusive identification occurs, it is guaranteed to be correct. 

In conventional quantum state discrimination  tasks, positive operator-valued measurements (POVMs) are typically employed as the standard measurement framework. In this work, we explore whether enhanced performance in state discrimination can be achieved by relaxing the positivity constraint of the measurement.
It is important to note that there exist information-processing tasks within quantum theory where non-positive operator-valued measurements (NPOVMs) outperform standard POVMs. Such instances of performance enhancement have been demonstrated in the context of generalized probabilistic theories in Refs.~\cite{HYM2019, YHM2020, MM2020, RSES+++23}. While this might raise concerns about the physical realizability of NPOVMs, Ref.~\cite{HM2025} provides a constructive framework for implementing NPOVMs using standard POVM measurements within the generalized probabilistic theories framework. There are several distinct contexts in which NPOVMs arise in the study of quantum state distinguishability. For instance, in~\cite{HM2025}, an NPOVM is defined as a collection of Hermitian operators 
$N := \{N_i\}_{i \in \mathcal{I}}$ satisfying the completeness condition
$\sum_{i \in \mathcal{I}} N_i = \mathbb{I}$,
where \( \mathbb{I} \) denotes the identity operator on the Hilbert space \( \mathcal{H} \), but where at least one of the elements \( N_i \) is not positive semi-definite. 
While such generalized measurement models are relevant in certain contexts, we do not explicitly adopt this particular definition in the present work.


The relationship between positive and non-positive operator-valued measurements is analogous to that between completely positive (CP) and non-completely positive (NCP) maps~\cite{AKU2024,anindita,PAKU2025}. In the conventional setting, NPOVMs emerge when a system is pre-entangled with an auxiliary system~\cite{PAKU2025,landauer}. A joint projective measurement is performed on the entangled state, and the auxiliary part is subsequently traced out. Mathematically, this leads to a non-positive effective measurement on the local system, depending on the initial entanglement and the joint measurement performed. A similar mechanism underlies the appearance of NCP maps in quantum operations which can be effectively implemented by entangling the system with an environment, applying a global unitary on the composite entangled state of the system and environment, and subsequently tracing out the environment, yielding a non-positive trace-preserving map on the system.
Interestingly, when considering QSD tasks, we encounter a scenario where NPOVMs emerge even without any pre-existing entanglement between auxiliary and system of interest. That is, a local NPOVM can arise purely from the structure of the discrimination problem itself, independent of conventional entanglement-based constructions. So this is in a sense another way in which  nonlocality without entanglement ~\cite{CDC++1999,MAUK2003} appears in quantum information theory.


\textit{Our task.} We start by asking a simple question, Is it possible to attain sub-helstrom bound? In the quest of answering this, we observe that the minimum error probability for discriminating between two quantum states $\rho_{B}$ and $\sigma_{B}$ can be reduced below the conventional Helstrom bound by extending the discrimination task to suitably constructed joint states $\rho_{AB}$ and $\sigma_{AB}$. Specifically, this reduction can be achieved either by introducing sufficient local indistinguishability between the auxiliary states $\rho_{A}$ and $\sigma_{A}$, or by establishing a degree of entanglement between the system $B$ and the auxiliary subsystem $A$, or by a combination of both effects.
 This reduction in error probability indicates the use of quantum measurement strategies on the subsytem that go beyond standard POVMs, which we refer to as NPOVMs.
Therefore, to motivate our approach, we find instances in quantum state discrimination tasks where the notion of such NPOVMs naturally appears.
To account for practical limitations, we analyze scenarios in which the available resources - namely, the local indistinguishability of the auxiliary states or the entanglement of the extended bipartite states - are present but constrained.
The central task reduces to identifying the optimal joint state, defined over the combined system and auxiliary subsystems, that maximizes the distinguishability of the target states under the imposed constraints. \\

The remainder of the paper is organized as follows. In Sec.~\ref{sec:Preliminaries}, we mention the preliminary definitions and setup required to 
formulate the problem. In ~\ref{sec:Results}, we 
present our main results. We then draw the conclusions of our work in Sec.~\ref{sec:Conclusion}. 

\section{Preliminaries}\label{sec:Preliminaries}
In this section, we provide a brief overview of the theoretical tools and relevant literature that form the basis of our analysis. \\

\textit{Trace norm}.- The trace norm between two states is a metric on the space of density operators and provides a measure of distinguishability between the two states. Let $\mathcal{D}(\mathcal{H})$ denote the set of density matrices on a Hilbert space $\mathcal{H}$. For any two quantum states $\rho, \sigma \in \mathcal{D}(\mathcal{H})$, the trace norm between them is defined as
\begin{equation}\label{eq:eqn1}
    ||\rho - \sigma||_1 := \mathrm{Tr}\left|\rho - \sigma\right|,
\end{equation}
where $|\rho - \sigma| = \sqrt{(\rho - \sigma)^\dagger (\rho - \sigma)}$, i.e. the sum of absolute values of the eigenvalues, of the square matrix $(\rho - \sigma)$ ~\cite{BC2009}. Trace distance is then defined as $\frac{1}{2} ||\rho - \sigma||_1$. \\

 
  \textit{Concurrence}.- For a pure two-qubit entangled state \( \ket{\psi} \), \textit{concurrence} is defined as
\begin{equation}\label{eq:eqn2}
    C(\ket{\psi}) := \left| \braket{\psi | \tilde{\psi}} \right|,
\end{equation}
where the spin-flipped state \( \ket{\tilde{\psi}} \) is given by
\[
\ket{\tilde{\psi}} = (\sigma_y \otimes \sigma_y) \ket{\psi}^*,
\]
with \( \ket{\psi}^* \) denoting the complex conjugate of \( \ket{\psi} \) in the computational basis, and \( \sigma_y \) being the Pauli-$y$ matrix~\cite{W98}. \\
 
\textit{Binary quantum state discrimination and the Helstrom bound}~\cite{C1969,A1973}.- 
Consider a scenario where Alice possess a classical random variable $X \in \{0, 1\}$, which can be thought of as the outcome of a biased coin flip, occurring with probabilities \(p_0\) and \(p_1\), respectively. Based on the outcome of this random variable, Alice prepares a quantum system \(Q\) in one of two quantum states: she prepares \(\rho\) if \(X = 0\), and \(\sigma\) if \(X = 1\). The resulting quantum state is then sent to Bob. Bob is aware of the two possible states \(\rho\) and \(\sigma\), as well as their prior probabilities, but he does not know which specific state was prepared in a given instance. His objective is to determine, with the highest possible accuracy, whether the received state is \(\rho\) or \(\sigma\). To this end, Bob is allowed to perform a general quantum measurement described by a positive-operator valued measure (POVM) \(\{M_i\}\), where the measurement operators satisfy the conditions \(M_i \geq 0\) for all \(i\), and \(\sum_i M_i = I\), with \(I\) being the identity operator acting on the Hilbert space $\mathcal{H}$.
If a measurement is performed by Bob with this set of POVM operators on the received state $\rho_a$, where $\rho_a \in \{\rho,\sigma\}$, then the probability that $M_i$ will be clicked is given by
\begin{eqnarray}
    q(i|a)=\Tr(\rho_a M_i),
\end{eqnarray}

The aim of Bob is to speculate on the received state with maximum probability. Upon obtaining an outcome corresponding to a POVM element \( M_i \), he guesses the prepared state to be \( \rho\) or \(\sigma\). A correct guess occurs when the index \( i \) associated with the measurement outcome matches the actual index \( a \) of the prepared state. 
For simplicity, in the remainder of the discussion we focus on the case where the two states are equally likely, i.e., \(p_0 = p_1 = \frac{1}{2}\). This setting captures the fundamental problem of binary quantum state discrimination, which lies at the heart of quantum hypothesis testing~\cite{C1969} and quantum information theory.
The average chances of success, i.e. correctly identifying the prepared state, when averaged over the entire ensemble, is given by
 \begin{eqnarray}
 \label{eq:eqn2}
 P_{\mathrm{success}} &=& p_0  \Tr(\rho M_0) + p_1  \Tr(\sigma M_1) \nonumber \\
 &=& \frac{1}{2} + \frac{1}{4} \operatorname{Tr}[(M_1 - M_2)(\rho - \sigma)] \nonumber  \\
 && \le \frac{1}{2} + \frac{1}{4} \| \rho - \sigma \|_1 
 \end{eqnarray}
The quantity, $P_{\mathrm{success}}$, is maximized when the POVM elements \( M_1 \) and \( M_2 \) are chosen as the projectors onto the positive and negative eigenspaces of the Hermitian operator \( \rho - \sigma \), respectively.  Hence, the maximum achievable success probability, known as the \textit{Helstrom bound}, is given by~\cite{C1969,A1973}
\begin{equation}\label{eq:eqn4}
P_{\mathrm{guess}} =  \max_{\{M_i\} \in \text{POVM}} P_{\mathrm{success}} = \frac{1}{2} + \frac{1}{4} \| \rho - \sigma \|_1 .
\end{equation}
The task of correctly identifying the prepared state from a given ensemble is the objective of \textit{binary quantum state discrimination}, and $P_{\mathrm{guess}}$ refers to the maximum success probability of guessing the prepared state.
Equivalently, we can say that the minimum probability of error, $P_{error}=1-P_{guess}$, in distinguishing between two quantum states $\rho$ and $\sigma$, given with equal \textit{a priori} probabilities, is characterized by the Helstrom error bound and is given by
\begin{equation}\label{eq:eqn6}
    P^{min}_{\text{err}} = \frac{1}{2} - \frac{1}{4} \left\| \rho - \sigma \right\|_1,
\end{equation}
where $\|\cdot\|_1$ denotes the trace norm.

It is natural at this stage to inquire whether the error probability in state discrimination tasks can be reduced below the Helstrom error bound by relaxing the requirement that measurement operators form a positive operator-valued measure (POVM). Achieving such a reduction would have profound implications for the theory and practice of quantum state discrimination, as the Helstrom bound is widely regarded as the fundamental limit under conventional measurement constraints. We now formalize a framework aimed at surpassing the Helstrom error bound.

Consider two bipartite quantum states, \(\rho_{AB}, \sigma_{AB} \in \mathcal{D}(\mathcal{H}_{AB})\), with no assumption on the nature or strength of the correlations shared between subsystems \(A\) and \(B\). We focus on the task of distinguishing the corresponding reduced states \(\rho_B = \operatorname{Tr}_A[\rho_{AB}]\) and \(\sigma_B = \operatorname{Tr}_A[\sigma_{AB}]\). 
In this setting, we pose the question whether it is possible to achieve an error probability in discriminating \(\rho_B\) and \(\sigma_B\) that is lower than the optimal limit set by the Helstrom bound. Specifically, we investigate whether such an improvement can be realized by having either of the following conditions or both, which are given by
\begin{itemize}
  \item imposing restrictions of local distinguishability on corresponding auxiliary subsystems of both the states $\rho_{AB}$ and $\sigma_{AB}$ i.e having partial  distinguishability between $\rho_{A}$ and $\sigma_{A}$.
    \item the presence of pre-shared entanglement between subsystems i.e auxiliary ($\rho_{A}$, $\sigma_{A}$) and system of interest ($\rho_{B}$, $\sigma_{B}$) 
    \end{itemize}

The objective of this investigation is to explore whether certain  local and global properties of the joint states, \(\rho_{AB}\) and \(\sigma_{AB}\), can 
improve the distinguishability of their respective marginal states in the subsystem \(B\), utilizing generalized measurements on subsystem, $B$ that includes non-positive ones. In other words, our objective is to determine whether such operational constraints can yield an error probability strictly below the standard Helstrom bound. 
From the Helstrom bound, it is known that the minimum probability of error in distinguishing between two quantum states \(\rho_{AB}\) and \(\sigma_{AB}\) is given by
\[
P_{\text{err}}^{\text{min}} = \frac{1}{2} - \frac{1}{4} \left\| \rho_{AB} - \sigma_{AB} \right\|_1,
\]
where \(\| \cdot \|_1\) denotes the trace norm. 

We here introduce a new minimum error probability to discriminate between the reduced states $\rho_{B}$ and $\sigma_{B}$. 
Given two quantum states $\rho_B$ and $\sigma_B$, each of which occurs with some known probability.
Alice prepares one of the states and sends it to Bob. Bob now discriminates among the two states by performing a generalized measurement.
To implement the generalized measurement, consider extending these states to their corresponding joint states $\rho_{AB}$ and $\sigma_{AB}$, which may be individually entangled, such that their marginals in the subsystem $B$ are $\rho_B$ and $\sigma_B$, respectively, and applying a joint projective measurement on the respective extended states,  $\rho_{AB}$ or $\sigma_{AB}$, followed by tracing out the auxiliary. 
In order to minimize the error probability, 
a minimization is performed on the parameters of the subsystem $A$, and the parameters of the joint projective measurement on the respective extended states. 
This essentially reduces the problem to distinguishing the two states, $\rho_{AB}$ and $\sigma_{AB}$, using the optimal projective measurement, which is  provided by the Helstrom bound, and an additional optimization over the parameters of subsystem, $A$. 
The aim is to find if the  probability of error in distinguishing $\rho_B$ and $\sigma_B$ can be reduced below the conventional Helstrom bound. 
However, without any constraints, this minimization can trivially yield an error probability of zero. To avoid such triviality and ensure a meaningful formulation, we impose resource constraints, i.e local non-orthogonality or global entanglement on the extension or both, which are incorporated as conditions in the optimization problem.
The optimization problem of estimating the minimum error in this scenario can be mathematically formulated as,
\begin{equation}
   \begin{aligned}
    P_{\text{error}}^{\text{NPOVM}} &= \min_{\rho_A,\sigma_A} \left( \frac{1}{2} - \frac{1}{4} \left\| \rho_{AB} - \sigma_{AB} \right\|_1 \right), \label{eq:eqn7} \\[6pt]
 \text{subject to} \quad
        &\begin{cases}
            \mathrm{Tr}_A(\rho_{AB}) = \rho_B, \\
            \mathrm{Tr}_A(\sigma_{AB}) = \sigma_B, \\
            D(\rho_A, \sigma_A) \le d, \\
            \max\{\mathcal{E}(\rho_{AB}), \mathcal{E}(\sigma_{AB})\} \le E.
        \end{cases}
        \end{aligned}
 \end{equation}
Here the minimization is performed over the parameters of the subsystem (auxiliary) \( A \), corresponding to both the joint states \( \rho_{AB} \) and \( \sigma_{AB} \). The functional \( D(\rho_A, \sigma_A) \) denotes a generic distance measure between the two states, $\rho_A $ and $ \sigma_A$. In this work, we specifically adopt the trace distance as the distance measure. The functional \( \mathcal{E}(\cdot) \) represents an entanglement measure, which we take to be the concurrence. The 
real numbers, \(d\) and \(E\), serve as bounds on the resources under consideration, namely the local distinguishability of subsystem \(A\) for both the states, and the joint entanglement shared between subsystems \(A\) and \(B\).

The aim is to find whether the identification of the optimal joint state that enables discrimination between $\rho_B$ and $\sigma_B$ can be obtained with an error probability strictly lower than that given by the conventional Helstrom bound.
If such a reduction in error probability is indeed achieved i.e $ P_{\text{error}}^{\text{NPOVM}} <  P_{\text{error}}^{\text{POVM}} $ where $P_{\text{error}}^{\text{POVM}}=\frac{1}{2} - \frac{1}{4} \left\| \rho_{B} - \sigma_{B} \right\|_1$ is the conventional helstrom bound to discriminate $\rho_B$ and $\sigma_B$ as per Eq.\eqref{eq:eqn6}, it will imply that the measurement employed 
does not form a POVM. Instead, they must involve \textit{non-positive operator-valued measurements (NPOVMs)}.
In the succeeding section, we present our results where we start with a simple example to get some intuition and then consider different possible cases under the described scenario.\\

\section{Binary quantum state discrimination using NPOVMs}\label{sec:Results}
To facilitate a 
better understanding 
of the general results, we begin by analyzing a simple illustrative example. This example highlights the fundamental aspects of the quantum state discrimination problem using non-positive operator-valued measurements (NPOVMs). 
Specifically, for this example, we do not assume any entanglement between subsystems $A$ and $B$ in the joint states $\rho_{AB}$ and $\sigma_{AB}$, thereby emphasizing the power of NPOVMs in scenarios devoid of entanglement.  For simplicity, we consider the states that we aim to discriminate
to be pure for this example. 
Nevertheless, to avoid trivial instances, we introduce local non-orthogonality between $\rho_A$ and $\sigma_A$.
Note that throughout our work, we only focus on qubit systems. Qubit case suffices as long as one is dealing with QSD task for just two states.
\begin{figure}
    \centering
\includegraphics[width=1.0\linewidth]{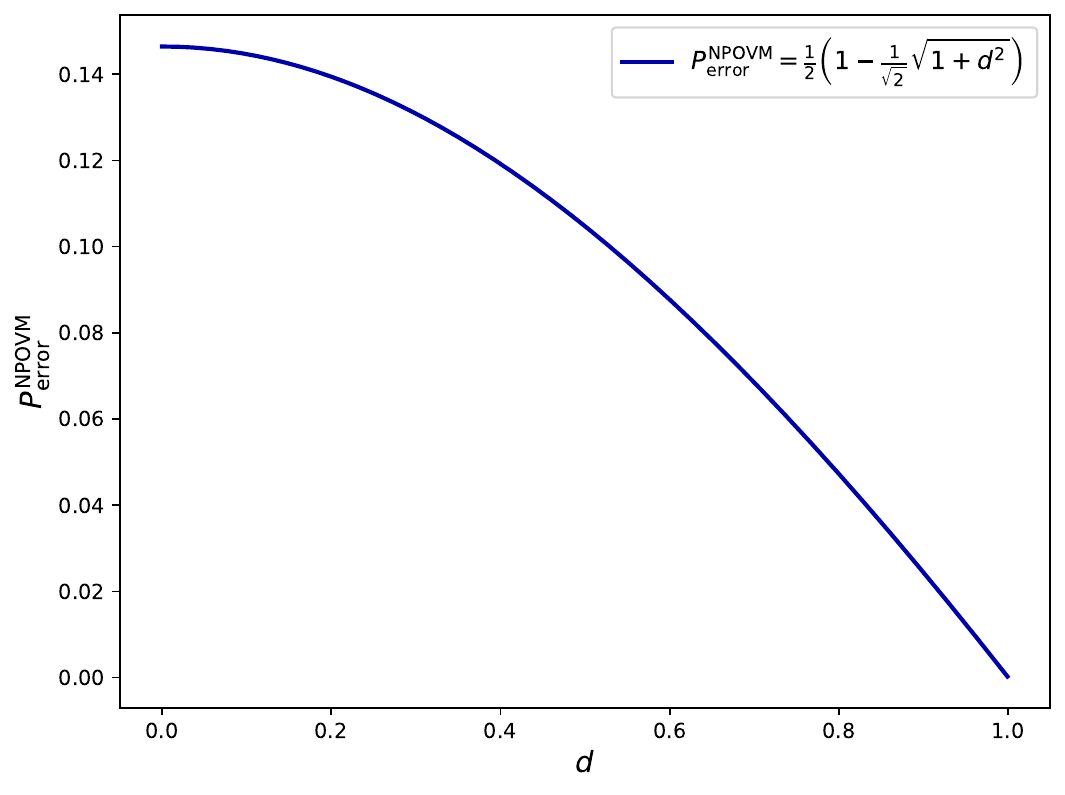}
    \caption{The minimum error probability, $P_{\text{error}}^{\text{NPOVM}}$, obtained if Bob pertains to generalized measurements on the received state is plotted along the vertical axes, versus the upper bound, $d$, on the distinguishability between the two states, $\rho_A$ and $\sigma_A$, which is plotted along the horizontal axes. The states which we aim to discriminate are $\rho_B=\ket{0}\bra{0}$ and $\sigma_B=\ket{+}\bra{+}$. The minimum error decreases monotonically with an increase in the value of $d$. The quantities plotted along both the axes are dimensionless.}
    \label{fig:example}
\end{figure}
\begin{example}\label{example1}
Consider the task of distinguishing between the two quantum states \(\rho_B = \ket{0}\bra{0}\) and \(\sigma_B = \ket{+}\bra{+}\), which are assumed to arise as reduced states of corresponding joint states \(\rho_{AB}\) and \(\sigma_{AB}\), respectively. Given that both \(\rho_B\) and \(\sigma_B\) are pure, it follows that the global states \(\rho_{AB}\) and \(\sigma_{AB}\) must be product states, up to local unitary transformations on the auxiliary system, since any purification of a pure state must necessarily be a product state (auxiliary and system). Moreover, even if the auxiliary system were initially in a mixed state, it can always be purified without loss of generality. Thus, for the purposes of this analysis, the joint states \(\rho_{AB}\) and \(\sigma_{AB}\) may be taken to be pure product states.
Therefore, we may assume $\rho_{AB}=\ket{\Psi}\bra{\Psi}_{AB}$ with $\ket{\Psi}_{AB}=\ket{\psi}_{A}\otimes \ket{0}_{B}$ and similarly, $\sigma_{AB}=\ket{\Phi}\bra{\Phi}_{AB}$ with $\ket{\Phi}_{AB}=\ket{\phi}_{A}\otimes \ket{+}_{B}$. Clearly, there is no entanglement present in this example. Therefore, in accordance with Eq.~\eqref{eq:eqn7}, the optimization problem in this case is given by
  \begin{equation}
    \begin{aligned}
        P_{\text{error}}^{\text{NPOVM}} &= \min \left( \frac{1}{2} - \frac{1}{4} \left\| \rho_{AB} - \sigma_{AB} \right\|_1 \right), \label{eq:eqn8} \\[6pt]
        \text{subject to} \quad & \left\{
            \begin{aligned}
                \mathrm{Tr}_A(\rho_{AB}) &= \rho_B \\
                \mathrm{Tr}_B(\rho_{AB}) &= \rho_A \\
                D(\rho_A, \sigma_A) &\le d 
            \end{aligned}
        \right.
    \end{aligned}
\end{equation}
where minimisation is with respect to auxiliary parameters. The entanglement constraint on the reduced states is redundant due to the specific form of the global states considered. Consequently, in this particular example, the only essential constraint have relevance to the local distinguishability of subsystem \(A\).

Let us consider the following choice of pure states for the subsystem \( A \), given by
\begin{align}
  \ket{\psi}_{A} &= \cos \theta \ket{0}_{A} + \sin \theta \ket{1}_{A}, \label{eq:eqn9}\\
  \ket{\phi}_{A} &= \cos \phi \ket{0}_{A} + \sin \phi \ket{1}_{A}, \label{eq:eqn10}
\end{align}
where \( \theta, \phi \in [0, \pi/2] \) parameterize the states on the Bloch sphere.
The corresponding density operators are given by \( \rho_{A} = \ket{\psi}\bra{\psi}_{A} \) and \( \sigma_{A} = \ket{\phi}\bra{\phi}_{A} \). The trace distance between these two pure states is 
\begin{align}
D(\rho_A, \sigma_A) = \sqrt{1 - |\braket{\psi|\phi}|^2} = |\sin(\theta - \phi)|. \label{eq:eqn11}
\end{align}
This expression captures the distinguishability between the two states, $\rho_A$ and $\sigma_A$. The trace distance attains its maximum value of 1 when the states are orthogonal (\( \theta - \phi = \pi/2 \)) and vanishes when the states are identical (\( \theta = \phi \)).
Therefore, the constraint \( D(\rho_A, \sigma_A) \leq d \) translates to the inequality \( |\sin(\theta - \phi)| \leq d \). The eigenvalue spectrum of matrix \(\rho_{AB} - \sigma_{AB}\) is explicitly given by:
\[
\left\{0, 0, -\frac{1}{2} \sqrt{3 - \cos\left[2(\theta - \phi)\right]}, \frac{1}{2} \sqrt{3 - \cos\left[2(\theta - \phi)\right]} \right\}
\] Recalling that the trace norm, as defined in Eq.~\eqref{eq:eqn1}, the trace norm between the joint states \(\rho_{AB}\) and \(\sigma_{AB}\) is as follows:
\begin{align}
    \| \rho_{AB} - \sigma_{AB} \|_1 = \sum_i |\lambda_i| = \sqrt{3 - \cos\left[2(\theta - \phi)\right]}\label{eq:eqn12}
\end{align}
where \(\lambda_i\) denote the eigenvalues of the Hermitian matrix \(\rho_{AB} - \sigma_{AB}\). 
Hence for this example, the minimum error probability, as defined in Eq.~\eqref{eq:eqn8}, takes the following form
   \begin{align}
P_{\mathrm{error}}^{\mathrm{NPOVM}}
&=\min_{\substack{\theta,\phi \\ |\sin(\theta-\phi)| \le d}}\left(
\frac{1}{2}-
\frac{1}{4}\sqrt{3 - \cos 2(\theta - \phi)}\right)
\\
&=\min_{\substack{\theta,\phi \\ |\sin(\theta-\phi)| \le d}}\left(
\frac{1}{2}-\frac{1}{2\sqrt{2}}
\sqrt{1 + \sin^2(\theta - \phi)}\right)
\end{align}
It is evident from the form of the objective function that the error probability is minimized when $|\sin(\theta - \phi)| = d$. Therefore, the optimal value is attained at the boundary of the constraint, with the minimum error being given by
\begin{equation}\label{eq:eqn15}
P_{\text{error}}^{\text{NPOVM}}  = \frac{1}{2} \left(1 - \frac{1}{\sqrt{2}} \sqrt{1 + d^2} \right).
\end{equation}
The following observation illustrates the behavior of the error probability as a function of the parameter $d$.
   For $d = 0$ i.e $\theta=\phi$, the error probability is $0.146$, consistent with the Helstrom bound for the given states i.e $\ket{0}\bra{0}$ and $\ket{+}\bra{+}$ where POVM's are employed.
For any $0 < d < 1$, the error probability achievable via our optimisation problem is strictly less than that attainable using standard POVMs. Note that $d=1$ makes the problem trivial, therefore for practicality we would restrict \textit{d} to be nonzero always but less than 1. These results highlight a distinct quantitative advantage offered by non-positive operator-valued measurements (NPOVMs), i.e. a strictly lower error probability for all $d > 0$. Specifically, we have
\[
P_{\text{error}}^{\text{NPOVM}} < P_{\text{error}}^{\text{POVM}} \quad \text{for all } d > 0.
\]
This inequality clearly demonstrates that extending beyond the conventional POVM framework can yield improved performance in quantum state discrimination tasks involving non-orthogonal states.
Fig.~\ref{fig:example} illustrates the variation of the error probability as a function of the distinguishability parameter $d$. As evident from the plot, the error probability decreases monotonically with increasing $d$, highlighting the advantage of employing generalized measurements specifically, non-positive operator-valued measurements (NPOVMs) over standard POVMs in the regime of restricted non-orthogonality between auxiliary states.
\end{example}

To quantify the distinction in performance between the two measurement strategies, we define a figure of merit as the difference in error probabilities between NPOVM and POVM strategies, given by
\[
\Delta P_{\text{er}} = P_{\text{error}}^{\text{NPOVM}} - P_{\text{error}}^{\text{POVM}}.
\]
A negative value of \( \Delta P_{\text{er}} \) indicates a clear operational advantage of NPOVMs over standard POVMs in distinguishing the quantum states available to Bob.
\\

We now proceed to generalize the preceding example through a sequence of increasingly general cases. In \textit{Case I}, we restrict our attention to product states, while allowing the pure states on subsystem $B$ to vary arbitrarily, subject to the condition that they remain non-orthogonal while having some restricted local distinguishability between subsystems A as was in the example discussed above. This serves as the first step toward a broader framework, which we further extend in the subsequent cases.\\

\subsection{Case I: Discriminating two arbitrary pure qubit states}
In this subsection, our objective is to discriminate between two arbitrary pure states $\ket{\chi}_{B}$ and $\ket{\delta}_{B}$ defined on subsystem $B$. 
  
  \begin{theorem}\label{th:1}
   For all pairs of pure  states, NPOVM's provide better quantum state discrimination than POVM's, for arbitrarily small distance between the auxiliary states, and without any system-auxiliary initial entanglement.
   \end{theorem}

\begin{figure}[h]
  \centering
  \includegraphics[width=\linewidth]{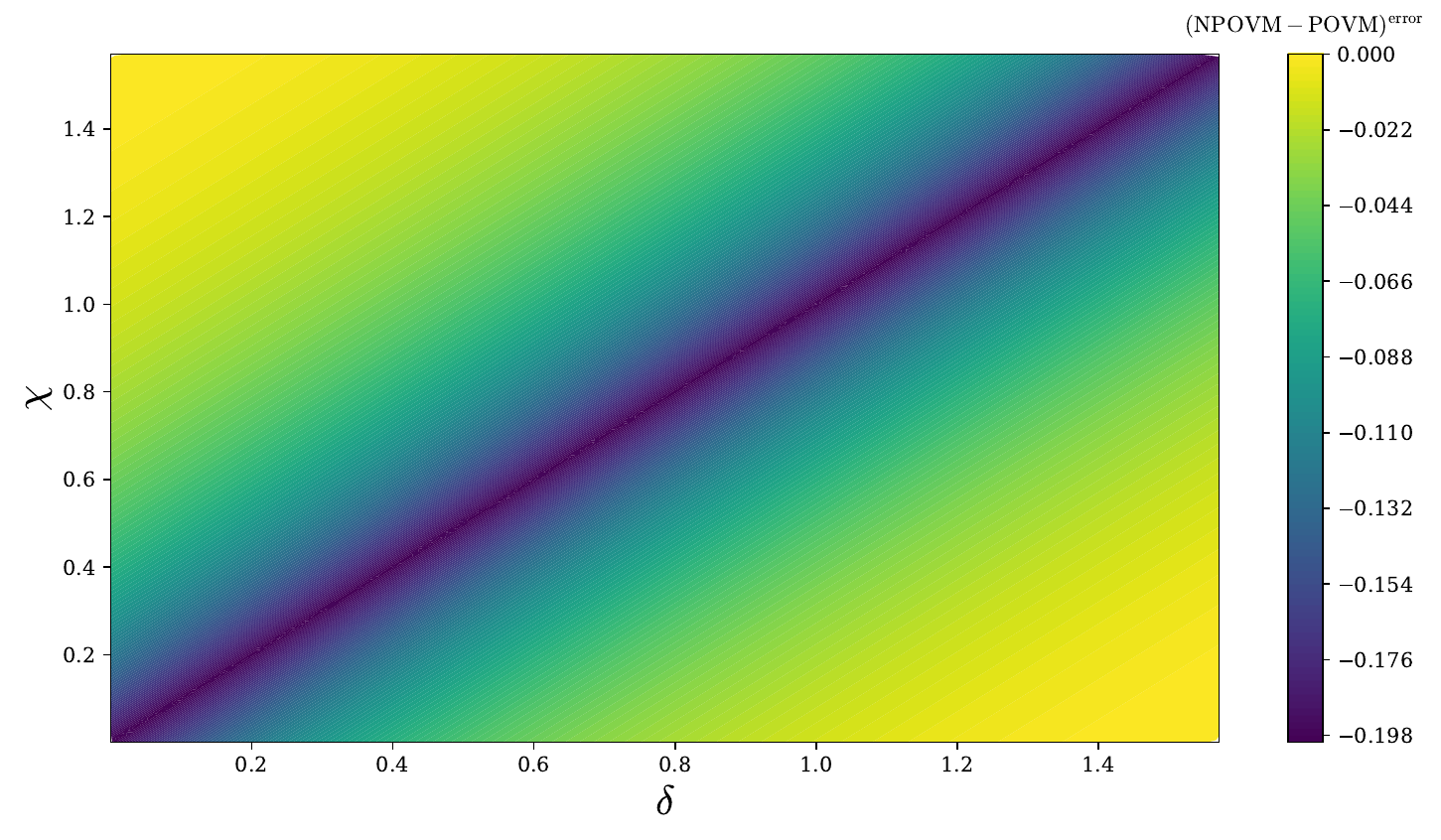}
  \caption{The difference in error probabilities between the NPOVM and standard POVM strategies, defined as \( \Delta P_{\text{er}} = P_{\text{error}}^{\text{NPOVM}} - P_{\text{error}}^{\text{POVM}} \), under the constraint that the distinguishability between the local states \( \rho_A \) and \( \sigma_A \) does not exceed a fixed bound \( d \). The scenario considered corresponds to Case I. The value of $d$ chosen is $0.4$. The overall error probability is minimized with respect to the parameters of subsystem \( A \). Notably, regions where \( \Delta P_{\text{er}} < 0 \) indicate an operational advantage of NPOVMs, yielding a strictly lower error probability compared to what is achievable with standard POVMs. The quantities plotted along all the axes are dimensionless.
  }
  \label{fig:case2}
\end{figure}
  
\noindent \textit{Proof.}  We consider the following composite quantum states as extensions of the states, $\ket{\chi}_{B}$ and $\ket{\delta}_{B}$, on the bipartite system $AB$, given by
\begin{align}
\rho_{AB} &= \ket{\psi} \bra{\psi}_{A} \otimes \ket{\chi} \bra{\chi}_{B}, \label{eq:eqn16}\\
\sigma_{AB} &= \ket{\phi} \bra{\phi}_{A} \otimes \ket{\delta} \bra{\delta}_{B}\label{eq:eqn17},
\end{align}
where the states $\ket{\chi}_{B}$ and $\ket{\delta}_{B}$ are parameterized as
\begin{align}
\ket{\chi}_{B} &= \cos\chi \ket{0}_{B} + \sin\chi \ket{1}_{B},\label{eq:eqn18} \\
\ket{\delta}_{B} &= \cos\delta \ket{0}_{B} + \sin\delta \ket{1}_{B}.\label{eq:eqn19}
\end{align}
Here the parameters, \( \chi, \delta \in [0, \pi/2] \), parameterize the states on the Bloch sphere. The corresponding pure states on the subsystem $A$, $\ket{\psi}_{A}$ and $\ket{\phi}_{A}$, are chosen to be the same as in the previously discussed example, given by
\begin{align}
\ket{\psi}_{A} &= \cos\theta \ket{0}_{A} + \sin\theta \ket{1}_{A}, \label{eq:eqn20}\\
\ket{\phi}_{A} &= \cos\phi \ket{0}_{A} + \sin\phi \ket{1}_{A}.\label{eq:eqn21}
\end{align}
The aim is to investigate whether improved discrimination can be achieved between the pure states $\chi_B$ and $\delta_B$, or equivalently, whether it is possible to attain an error probability strictly lower than that predicted by the Helstrom bound for their discrimination. It is important to recall that in this case also, there is absence of any 
entanglement between the subsystems $A$ and $B$ in the corresponding states $\rho_{AB}$ and $\sigma_{AB}$. 
However, a constraint regarding subsystems A persists, specifically, the trace distance between the reduced states on the subsystems $A$ must satisfy
$D(\rho_A, \sigma_A) \leq d$,
which imposes a bound on the distinguishability between the two states, $\rho_{A}$ and $\sigma_{A}$. This constraint plays a central role in limiting the achievable performance in the discrimination task under consideration.

Recall our optimization problem, which is given by
\begin{equation}\label{eq:eqn22}
    \begin{aligned}
        P_{\text{error}}^{\text{NPOVM}} &= \min \left( \frac{1}{2} - \frac{1}{4} \left\| \rho_{AB} - \sigma_{AB} \right\|_1 \right), \\
        \text{subject to} \quad & \left\{
            \begin{aligned}
                \mathrm{Tr}_A(\rho_{AB}) &= \rho_B \\
                \mathrm{Tr}_B(\rho_{AB}) &= \rho_A \\
                D(\rho_A, \sigma_A) &\le d.
            \end{aligned}
        \right.
    \end{aligned}
\end{equation}
Evaluating the trace norm between the states $\rho_{AB}$ and $\sigma_{AB}$, followed by a few steps of simplification, gives,
\begin{align}
 &\left( \frac{1}{2} - \frac{1}{4} \left\| \rho_{AB} - \sigma_{AB} \right\|_1 \right) \nonumber \\
&=\frac{1}{2}-\frac{1}{2} \sqrt{\sin^2(\theta - \phi) \cos^2(\delta - \chi)+ \sin^2(\delta - \chi}). \label{eq:eqn24}
\end{align}
%
The trace distance between the reduced states $\rho_{A}$ and $\sigma_{A}$, as defined by Eq.~\eqref{eq:eqn1}, is 
given by $D(\rho_{A}, \sigma_{A})=| \sin(\theta - \phi)|$.
So, finally, the optimization problem is given by the following
\begin{align}
\min_{\substack{\theta,\phi \\ |\sin(\theta-\phi)| \le d}} \left( \frac{1}{2} - \frac{1}{2} 
\sqrt{
\sin^2(\theta - \phi) \cos^2(\delta - \chi) + \sin^2(\delta - \chi)
}
\right) \label{eq:eqn25}
\end{align}

\noindent
After incorporating the constraint into the optimization problem, we find
that the error probability under our constrained optimisation is given by
\begin{equation}
P_{\text{error}}^{\text{NPOVM}} = \frac{1}{2} - \frac{1}{2} \sqrt{d^2 \cos^2(\delta - \chi) + \sin^2(\delta - \chi)} \label{eq:eqn26}.
\end{equation}
For the special case $\chi = 0$ and $\delta = \pi/4$, the objective function yields
\begin{equation}\label{eq:eqn27}
P_{\text{error}}^{\text{NPOVM}} = \frac{1}{2} \left(1 - \frac{1}{\sqrt{2}} \sqrt{1 + d^2} \right),
\end{equation}
which coincides with the result obtained in the example. The condition, $d=0$, provides the minimum error probability using POVM. So, from the above equation, we find that for any arbitrarily small but nonzero $d > 0$, we have
\[
P_{\text{error}}^{\text{NPOVM}} < P_{\text{error}}^{\text{POVM}},
\]
demonstrating a consistent advantage offered by NPOVMs in the state discrimination task. 
For completeness, in Fig.~\ref{fig:case2}, we illustrate the difference in error probabilities between the NPOVM and POVM cases, i.e. $\Delta P_{er}=P_{\text{error}}^{\text{NPOVM}}-P_{\text{error}}^{\text{POVM}}$, obtained under the scenario in which the distinguishability between the local states, $\rho_A$ and $\sigma_A$, is upper bounded by a particular value of  $d$, and a minimization of the overall error probability is performed over the parameters of the subsystem, $A$. In particular, 
for the entire ranges of $\chi$ and $\delta$, the figure of merit, $\Delta P_{er}$ gives a negative value, suggesting that
the error obtained using NPOVMs is strictly lower than that achievable via standard POVMs, i.e., $P_{\text{error}}^{\text{NPOVM}} < P_{\text{error}}^{\text{POVM}}$. \hfill $\blacksquare$
%
%

\vspace{0.2cm}


Up to this point, our analysis has been restricted to the discrimination of pure quantum states. We now turn our attention to the more general scenario in which the states to be discriminated are mixed qubit states. \\

\subsection{Case II:  Discriminating two arbitrary single-qubit states}
We here generalize the theorem~\ref{th:1} for a pair of single-qubit pure states to the set of single-qubit arbitrary states, including mixed ones.

 \begin{theorem}\label{th:2}
  For all pairs of single-qubit states, including mixed ones that are prepared by Alice, and one of the states is sent to Bob, it is found that NPOVMs implemented at Bob's end yield a lower minimum error probability for distinguishing the two states compared to POVMs.
  The advantage using NPOVM persists even when the auxiliary states are arbitrarily close and in the absence of any initial entanglement between the system and the auxiliary. 
\end{theorem}

\noindent \textit{Proof.}  In this case, let the joint states be of the form 
\begin{align*}
\rho_{AB} &= \ket{\psi}\bra{\psi}_{A} \otimes \rho_B \;\; \text{and} \\
\sigma_{AB} &= \ket{\phi}\bra{\phi}_{A} \otimes \sigma_B,
\end{align*}
where $\rho_B$ and $\sigma_B$ denote the arbitrary quantum states in $\mathbb{C}^2$, which are to be discriminated, while the auxiliary subsystems remain in pure states. It is evident that there is no 
entanglement present in the joint states. Nevertheless, we allow the auxiliary subsystems to exhibit a finite degree of restricted local distinguishability, in analogy with the earlier formulation. 
The marginal states on subsystem \(B\) are considered to be arbitrary states given as,
\[
\rho_B = \frac{\mathbb{I} + \vec{m} \cdot \vec{\sigma}}{2}, \;\; \text{and} \quad 
\sigma_B = \frac{\mathbb{I} + \vec{n} \cdot \vec{\sigma}}{2}.
\]
Here, \(\{\vec{m}, \vec{n} \}\in \mathbb{R}^3\) are Bloch vectors, and \(\vec{\sigma} = (\sigma_x, \sigma_y, \sigma_z)\) denotes the vector of Pauli matrices. The states \(\ket{\psi}_A\) and \(\ket{\phi}_A\) correspond to those considered in the previous 
theorem, and characterize the reduced states on subsystem \(A\).
We again consider  restriction on the degree of local distinguishability for the reduced states on subsystem \(A\). Specifically, the trace distance between \(\rho_A\) and \(\sigma_A\) is constrained as  
$D(\rho_A, \sigma_A) \leq d$ 
which, for the considered pure states, translates to the condition \(|\sin(\theta - \phi)| \leq d\). Also,
\[
\left\lVert \rho_B - \sigma_B \right\rVert_1 = \left| \vec{m} - \vec{n} \right|,
\]  
where \(\vec{m}\) and \(\vec{n}\) denote the respective Bloch vectors of \(\rho_B\) and \(\sigma_B\).
Using subadditivity of trace norm under tensor products we have,
\begin{align*}
||{\rho_{AB} - \sigma_{AB}}||_{1} &= ||{ \ket{\psi}\bra{\psi} \otimes \rho_B - \ket{\phi}\bra{\phi} \otimes \sigma_B }||_{1} \\
&\leq ||{ \ket{\psi}\bra{\psi} - \ket{\phi}\bra{\phi} }||_{1} + ||{ \rho_B - \sigma_B }||_{1} \\
&=  |{\sin(\theta - \phi)}| + ||{ \rho_B - \sigma_B }||_{1},
\end{align*}
which implies $
\left\lVert \rho_{AB} - \sigma_{AB}\right\rVert_1 
\leq  |{\sin(\theta - \phi)}| + |{\vec{m} - \vec{n}}|$.
Therefore, the optimization problem, given by 
\begin{equation}\label{eq:eqn28}
    \begin{aligned}
        P_{\text{error}}^{\text{NPOVM}} &= \min_{\substack{\theta,\phi \\ |\sin(\theta-\phi)| \le d}} \left( \frac{1}{2} - \frac{1}{4} \left\| \rho_{AB} - \sigma_{AB} \right\|_1 \right) 
    \end{aligned}
\end{equation}
in this case takes the following form,
\begin{align}
P_{\text{error}}^{\text{NPOVM}} 
&\geq \min_{\substack{|\sin(\theta-\phi)| \le d}} \left( \frac{1}{2} - \frac{1}{4} \left[ \left| \sin(\theta - \phi) \right| + \left| \vec{m} - \vec{n} \right| \right] \right) \notag \\
&= \frac{1}{2} - \frac{1}{4} \max_{\, \left| \sin(\theta - \phi) \right| \leq d} \left( \left| \sin(\theta - \phi) \right| + \left| \vec{m} - \vec{n} \right| \right) \notag \\
&= \frac{1}{2} - \frac{1}{4} \left( d + \left| \vec{m} - \vec{n} \right| \right). \label{eq:eqn29}
\end{align}
Note that inequality~\eqref{eq:eqn29} provides a lower bound of the minimum error probability obtained using NPOVM. Whether the bound is tight is yet to be proven analytically.
Similarly to the previous case, here, the condition $d=0$, also provides the limit of the minimum error probability using POVM, i.e. $P_{\text{error}}^{\text{POVM}}$. Therefore, the above
inequality suggests that even in the absence of entanglement between the system and auxiliary, it is possible for any arbitrarily small \( d > 0 \) to discriminate between the mixed states \(\rho_B\) and \(\sigma_B\) with an error probability strictly below the conventional Holevo-Helstrom bound, assuming the bound is attainable. We further demonstrate numerically that this reduced error probability is indeed achievable.

\begin{figure}[h]
  \centering
  \includegraphics[width=\linewidth]{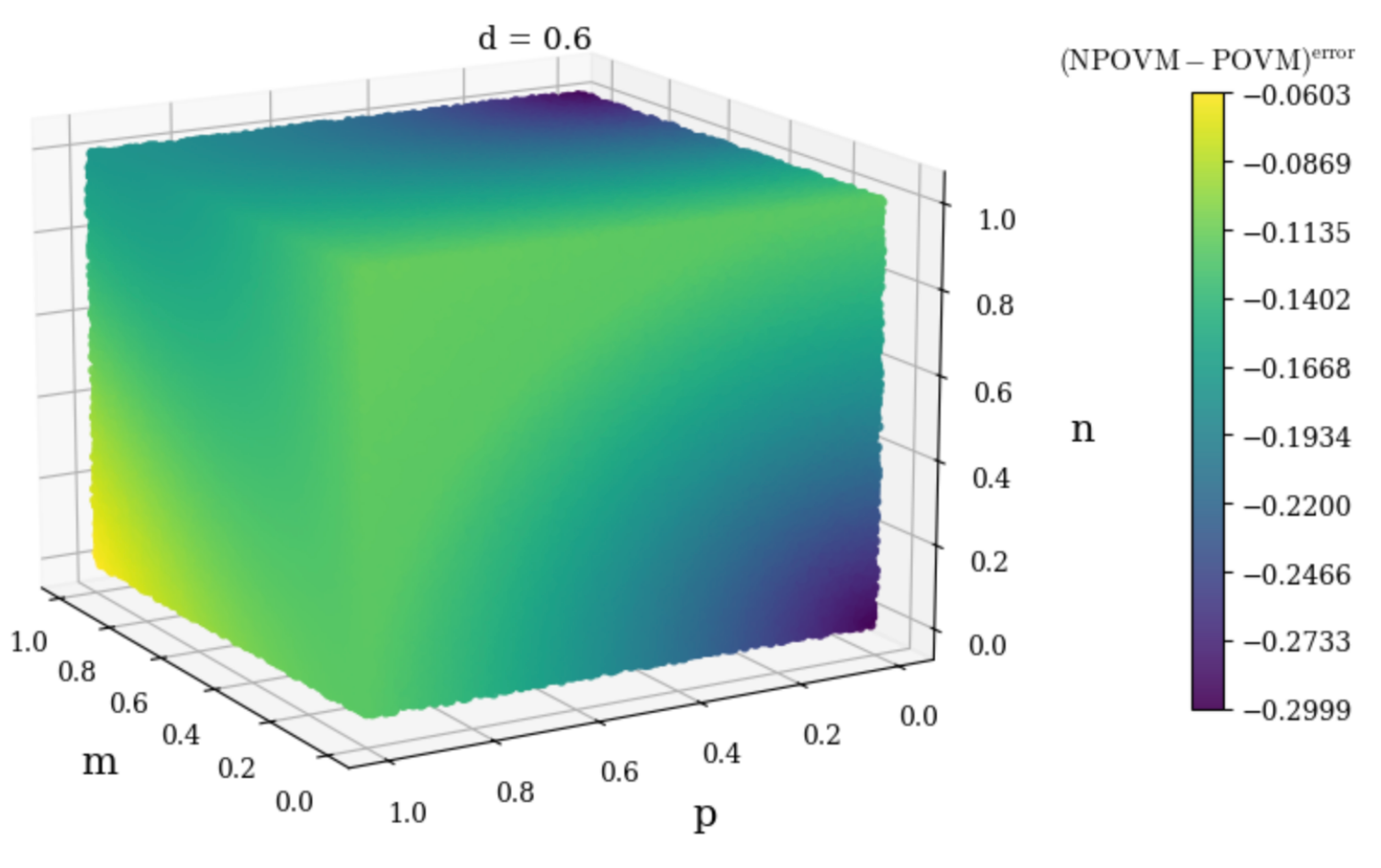}
  \caption{The figure of merit,  \(\Delta P_{er}=P_{\text{error}}^{\text{NPOVM}} - P_{\text{error}}^{\text{POVM}}\), plotted  as a function of the parameters of the state, \(n\), \(m\) and \(p\), illustrating the advantage of NPOVMs under local indistinguishability, under the condition that that the local distinguishablity of the subsystem, $A$ does not exceed a bound, $d$. Here, $d$ is chosen to be $0.6$. The analysis is conducted within the framework of Case II. The quantities plotted along all the axes are dimensionless.}
  \label{fig:ncase3}
\end{figure}

Since the two quantum states under consideration lie within the Bloch sphere, without loss of generality, they can be assumed to lie in a common plane. By applying an appropriate global unitary transformation, we can align the Bloch vector of \(\rho_B\) along the \(z\)-axis, such that \(\vec{m} = (0, 0, m)\), and represent the Bloch vector of \(\sigma_B\) within the \(x\)-\(z\) plane as \(\vec{n} = (p, 0, n)\). Fig.~\ref{fig:ncase3} illustrates the difference between the error probabilities obtained using NPOVM and POVM, i.e. $\Delta P_{er}=P_{\text{error}}^{\text{NPOVM}}-P_{\text{error}}^{\text{POVM}}$, while the values of the parameter, $d$ is fixed at $d=0.6$. The figure depicts that the for all values of $m$, $n$ and $p$, the figure of merit gives a negative value, implying an advantage of using NPOVM over POVM in discriminating the two states, $\rho_B$ and $\sigma_B$. \hfill $\blacksquare$ \\

In cases I and II, we found instances of implementing an NPOVM on a qubit, where the pure-state extensions of the qubit are trivial in the sense that they are still unentangled.
We now proceed to the next case, where we consider the presence of 
entanglement between the subsystems \(A\) and \(B\) in the joint states \(\rho_{AB}\) and \(\sigma_{AB}\). The inclusion of entanglement activates an additional constraint in the optimization problem defined in Eq.~(\ref{eq:eqn7}).

\subsection{Case III: Discriminating two arbitrary incoherent states when the pure-state extensions are entangled}
Here our aim is to distinguish between the two states, $\rho_{B}$ and $\sigma_{B}$, while considering their respective joint states, $\rho_{AB}$ and $\sigma_{AB}$, to individually have a certain amount of entanglement, along with the presence of partial distinguishability between the reduced states of subsystem, $A$.

\begin{theorem}
   There exist pairs of single-qubit incoherent states prepared by Alice, 
   whose purifications onto subspace $AB$ generate two-qubit pure entangled states, for which it is observed that NPOVMs implemented at Bob's end achieve a lower minimum error probability in state discrimination compared to standard POVMs. Interestingly, the advantage offered by NPOVMs persists even when the auxiliary states are arbitrarily close.
\end{theorem}

\noindent \textit{Proof.} Let \(\rho_{AB} = \ket{\Psi_{AB}}\bra{\Psi_{AB}}\) and \(\sigma_{AB} = \ket{\Phi_{AB}}\bra{\Phi_{AB}}\) be pure bipartite entangled states on the Hilbert space \(\mathcal{H}_A \otimes \mathcal{H}_B\), defined as follows
\begin{align}
  \ket{\Psi_{AB}} &= \sqrt{\lambda} \ket{\psi_A} \otimes \ket{0}_B + \sqrt{1 - \lambda} \ket{\psi_A^\perp} \otimes \ket{1}_B, \label{eq:eqn30}\\
  \ket{\Phi_{AB}} &= \sqrt{\mu} \ket{\phi_A} \otimes \ket{0}_B + \sqrt{1 - \mu} \ket{\phi_A^\perp} \otimes \ket{1}_B, \label{eq:eqn31}
\end{align}
where the variables \(\lambda,\mu \in (0,1)\) denote the Schmidt coefficients. 
Note that the states that we aim to discriminate, i.e. the reduced states of $\rho_{AB}$ and $\sigma_{AB}$ onto the subsystem \( B \) are incoherent in the computational basis, and are given by $\rho_B = \mathrm{Tr}_A(\rho_{AB}) = diag(\lambda, 1-\lambda)$ and $\sigma_B = \mathrm{Tr}_A(\sigma_{AB}) = diag(\mu, 1-\mu)$
respectively, where the notation $diag(.,.)$, denotes the diagonal elements of a $2\times 2$ matrix in the computational basis. 
Since we consider bipartite pure states in
which each subsystem is a qubit,
the Schmidt rank of the bipartite state is at most equal to $2$. Consequently, the above expression for the bipartite entangled state contains only two terms in its Schmidt decomposition. The local states \(\ket{\psi_A}, \ket{\psi_A^\perp}, \ket{\phi_A}, \ket{\phi_A^\perp} \in \mathcal{H}_A\) are defined through the following equations, given by
\begin{align}
  \ket{\psi_A} &= \cos\theta \ket{0} + \sin\theta \ket{1}, \label{eq:eqn32}\\
  \ket{\psi_A^\perp} &= \sin\theta \ket{0} - \cos\theta \ket{1}, \label{eq:eqn33} \\
  \ket{\phi_A} &= \cos\phi \ket{0} + \sin\phi \ket{1}, \label{eq:eqn34}\\
  \ket{\phi_A^\perp} &= \sin\phi \ket{0} - \cos\phi \ket{1}. \label{eq:eqn35}
\end{align}
Here, \(\ket{\psi_A^\perp}\) and \(\ket{\phi_A^\perp}\) denote the states orthogonal to \(\ket{\psi_A}\) and \(\ket{\phi_A}\), respectively. 
Note that the auxiliary states, i.e. the reduced states of $\rho_{AB}$ and $\sigma_{AB}$,  on the subsystem $A$, have 
non-zero coherence in the computational basis.

\begin{figure}
  \centering
  \includegraphics[width=\linewidth]{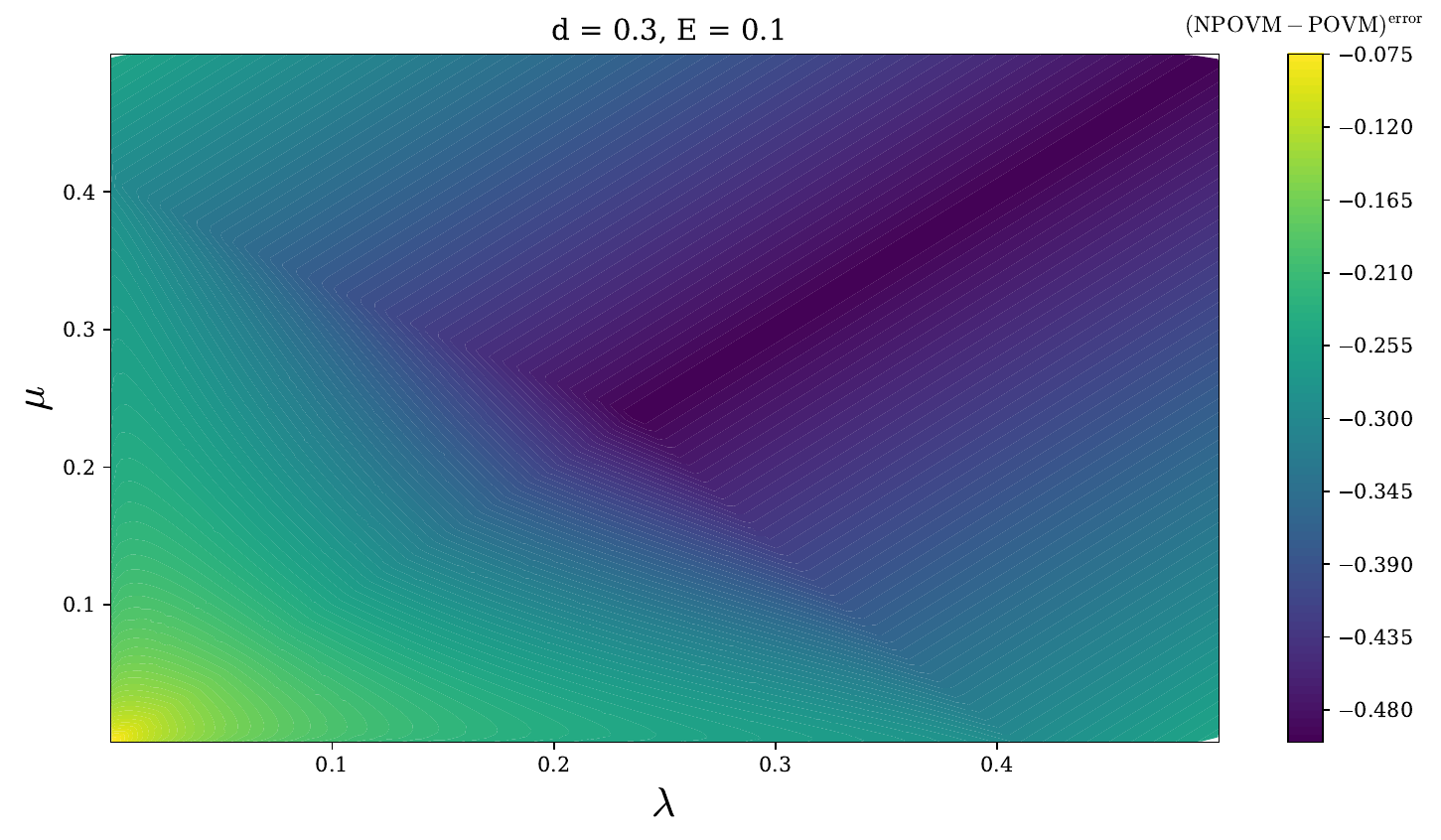}
  \caption{The figure of merit, \(\Delta P_{er}=P_{\text{error}}^{\text{NPOVM}} - P_{\text{error}}^{\text{POVM}}\), is plotted as a function of the state parameters, parameters \(\lambda\) and \(\mu\), illustrating the advantage of NPOVMs under restricted local indistinguishability ($d=0.3$) and global entanglement ($E=0.1$). The scenario considered corresponds to Case III. The quantities plotted along each of the axes are dimensionless.}
  \label{fig:ncase4}
\end{figure}

Next, we impose restrictions on the allowed resources, which gives the
constraints for our optimization problem. 
We use concurrence for pure two-qubit states~\cite{W98} as a measure of entanglement. 
The concurrence of a pure two-qubit state, $\ket{\psi}$, is defined by
\begin{align}
   C(\ket{\psi}) = |\braket{\psi | \tilde{\psi}}|,\label{eq:eqn36}
 \end{align}
where the state, $\ket{\tilde{\psi}} = \sigma_y \otimes \sigma_y \ket{\psi^*}$. Here $\ket{\psi^*}$ denotes the conjugate of the state, $\ket{\psi}$. 
In our case, the concurrences of the states, $\ket{\psi_{AB}}$ and $\ket{\phi_{AB}}$, are respectively given by 
\begin{align}
   C(\ket{\psi_{AB}})&=2\sqrt{\lambda(1-\lambda)}, \;\; \text{and}\label{eq:eqn37}\\
    C(\ket{\phi_{AB}})&=2\sqrt{\mu(1-\mu)}.\label{eq:eqn38}
\end{align}
Therefore, our optimization problem in this scenario can be mathematically stated in the following way,
\begin{equation}\label{eq:eqn39}
    \begin{aligned}
        P_{\text{error}}^{\text{NPOVM}} &= \min_{\substack{\theta,\phi  \\
        \lVert \rho_A - \sigma_A \rVert_1 \leq d\\
        \max \left\{ 2\sqrt{\lambda(1-\lambda)}, 2\sqrt{\mu(1-\mu)} \right\} \leq E}}\left( \frac{1}{2} - \frac{1}{4} \left\| \rho_{AB} - \sigma_{AB} \right\|_1 \right)
    \end{aligned}
\end{equation}
where $\rho_{AB}$ and $\sigma_{AB}$ are given by $\ket{\Psi_{AB}}\bra{\Psi_{AB}}$ and $\ket{\Phi_{AB}}\bra{\Phi_{AB}}$, respectively, and are defined in Eqs.~\eqref{eq:eqn30} and~\eqref{eq:eqn31}. While, $\rho_A$ and $\sigma_A$, are the reduced density matrix of the states $\rho_{AB}$ and $\sigma_{AB}$ onto the subsystem $A$. 

Fig.~\ref{fig:ncase4} illustrates the figure of merit,  $\Delta P_{er}=P_{\text{error}}^{\text{NPOVM}}-P_{\text{error}}^{\text{POVM}}$, as functions of the state parameters, $\mu$ and $\lambda$, for the values of the real numbers $d=0.3$ and $E=0.1$. We find that in the entire range of $\mu$ and $\lambda$, the figure of merit gives a negative value, implying an advantage of using non-positive measurements over POVMs even in the discrimination of two single-qubit states whose pure state-extensions are entangled, and have a non-zero indistinguishability among the auxiliary states. \hfill $\blacksquare$ \\

\subsection{Case IV: Discriminating two arbitrary qubits having finite coherence and entangled pure-state extensions}

\begin{theorem}
    There exist pairs of single-qubit states with non-zero coherence that are prepared by Alice whose purifications to the joint system \( AB \) yield pure entangled two-qubit states. For such configurations, it is observed that NPOVMs implemented at Bob's end result in a lower minimum error probability for state discrimination compared to standard POVMs. Notably, this advantage persists even when the auxiliary states are nearly indistinguishable.
    \label{new}
\end{theorem}

\noindent \textit{Proof.} We investigate the role of entanglement in the extended system state, along with the local distinguishability of the auxiliary states, in enhancing state discrimination through measurements that go beyond POVMs, while one of the states to be discriminated contain a non-zero coherence in computational basis.
To ensure this, we select the 
local densities of the subsystem $A$, to be diagonal in the computational basis, with a finite distinguishability between the auxiliary states, and in this case we particularly choose
$\ket{\phi_A} = \ket{\psi_A}$ and $\ket{\phi_A^\perp} = \ket{\psi_A^\perp}$, and $\lambda \ne \mu$ (refer to   Eqs.~\eqref{eq:eqn30} and~\eqref{eq:eqn31}). 
Note that the auxiliary states are indistinguishable in the quantum sense, and they are only classically distinguishable.
Moreover, in contrast to the preceding case, we do not consider the reduced states corresponding to the subsystem $B$ to be diagonal in the computational basis. 
The pure bipartite entangled states under this consideration are then given by
\begin{align}
\ket{\Psi_{AB}} &= \sqrt{\lambda} \ket{\psi_A} \otimes \ket{0} + \sqrt{1 - \lambda} \ket{\psi_A^\perp} \otimes \ket{1}, \;\; \text{and} \label{eq:eqn40}\\
\ket{\Phi_{AB}} &= \sqrt{\mu} \ket{\psi_A} \otimes \ket{0'} + \sqrt{1 - \mu} \ket{\psi_A^\perp} \otimes \ket{1'}. \label{eq:eqn41}
\end{align}
Here, the shared Schmidt basis vectors on the subsystem $A$ are parameterized by
\begin{align}
\ket{\psi_A} &= \cos\theta \ket{0} + \sin\theta \ket{1}, \;\; \text{and} \label{eq:eqn42}\\
\ket{\psi_A^\perp} &= \sin\theta \ket{0} - \cos\theta \ket{1}, \label{eq:eqn43}
\end{align}
where  \( \theta \in [0, \pi/2] \). Note that the basis states on the subsystem $B$ in the second state $\ket{\Phi_{AB}}$ are non-orthogonal, which are defined through the following relations,
\begin{align}
\ket{0'} &= \cos x \ket{0} + e^{i y} \sin x \ket{1},  \;\; \text{and} \label{eq:eqn44}\\
\ket{1'} &= \sin x \ket{0} - e^{i y} \cos x \ket{1}. \label{eq:eqn45}
\end{align}
where variables, \(x \in [0, \pi/2] \) and \(y \in [0, 2\pi] \).
The states which we aim to discriminate, \(\rho_{B}\) and \(\sigma_{B}\), expressed in the computational basis 
are respectively given by $\rho_B=diag(\lambda,1-\lambda)$, and
\begin{equation}\label{eq:eqn47}
\sigma_B = 
\begin{bmatrix}
c & a - ib \\
a + ib & 1 - c
\end{bmatrix}, 
\end{equation}
where the quantities, \(a, b, c \in \mathbb{R}\), are defined by 
$ a - ib = \sin 2x e^{-iy}(2\mu - 1)/2,$ and $c = \sin^2 x + \mu \cos 2x$.
\begin{figure}[h]
  \centering
  \includegraphics[width=\linewidth]{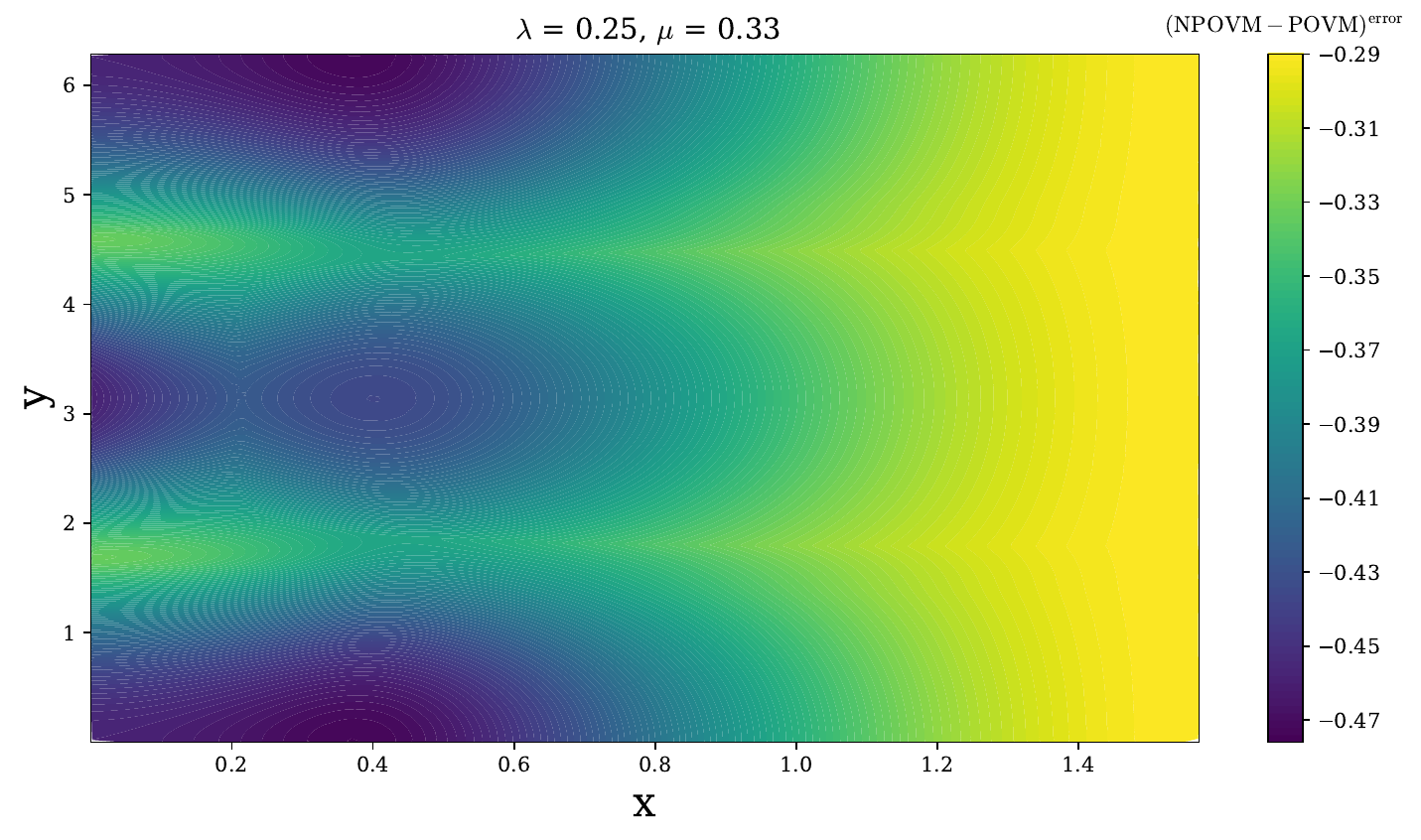}
  \caption{The figure of merit, defined as \( \Delta P_{\text{er}} = P_{\text{error}}^{\text{NPOVM}} - P_{\text{error}}^{\text{POVM}} \), is plotted as a function of the state parameters \( x \) and \( y \), to illustrate the operational advantage of NPOVMs over POVMs. The analysis is conducted corresponds to Case IV where the extended joint states are entangled. The plot corresponds to fixed values \( \lambda = 0.25 \) and \( \mu = 0.33 \). All quantities displayed along the axes are dimensionless.
  }
  \label{fig:case5}
\end{figure}
%
While both $\rho_B$ and $\sigma_B$ may, in general, exhibit coherence, one can always choose a basis in which either $\rho_B$ or $\sigma_B$ is diagonal. Without loss of generality, we assume this diagonal basis to be the computational basis, i.e. $\{\ket{0}, \ket{1}\}$. Consequently, it is sufficient to consider coherence in only one of the two  states. 

We now examine the constraints imposed in this setting, which are given by
\begin{itemize}
    \item[(i)] the condition on the distinguishability of the two auxiliary states, i.e. $D(\rho_A, \sigma_A) \leq d$, which reduces to $2|\lambda - \mu| \leq d$ in this case, and  
     \item[(ii)] the condition on the entanglement of the joint states, given by $\max\{E(\rho_{AB}), E(\sigma_{AB})\} \leq E$.
\end{itemize}
Here, $E(\cdot)$ is a suitable entanglement quantifier. Again consider the concurrence to be the measure of entanglement. For the pure states, $\ket{\Psi_{AB}}$ and $\ket{\Phi_{AB}}$, the concurrence are given by
\[
C(\ket{\Psi_{AB}}) = 2\sqrt{\lambda(1 - \lambda)}, \quad
C(\ket{\Phi_{AB}}) = 2\sqrt{\mu(1 - \mu)},
\]
where $\lambda$ and $\mu$ denote the Schmidt coefficients of $\ket{\psi_{AB}}$ and $\ket{\phi_{AB}}$, respectively. The entanglement constraint (ii) then implies
\[
\Rightarrow 
\begin{cases}
2\sqrt{\lambda(1 - \lambda)} \leq E, & \text{if } \lambda > \mu \\
2\sqrt{\mu(1 - \mu)} \leq E, & \text{if } \mu > \lambda
\end{cases}
\]
In the case where $\lambda = \mu$, both states possess equal entanglement, and the condition simplifies to $2\sqrt{\lambda(1 - \lambda)} \leq E$.

\begin{figure}
  \centering
  \includegraphics[width=\linewidth]{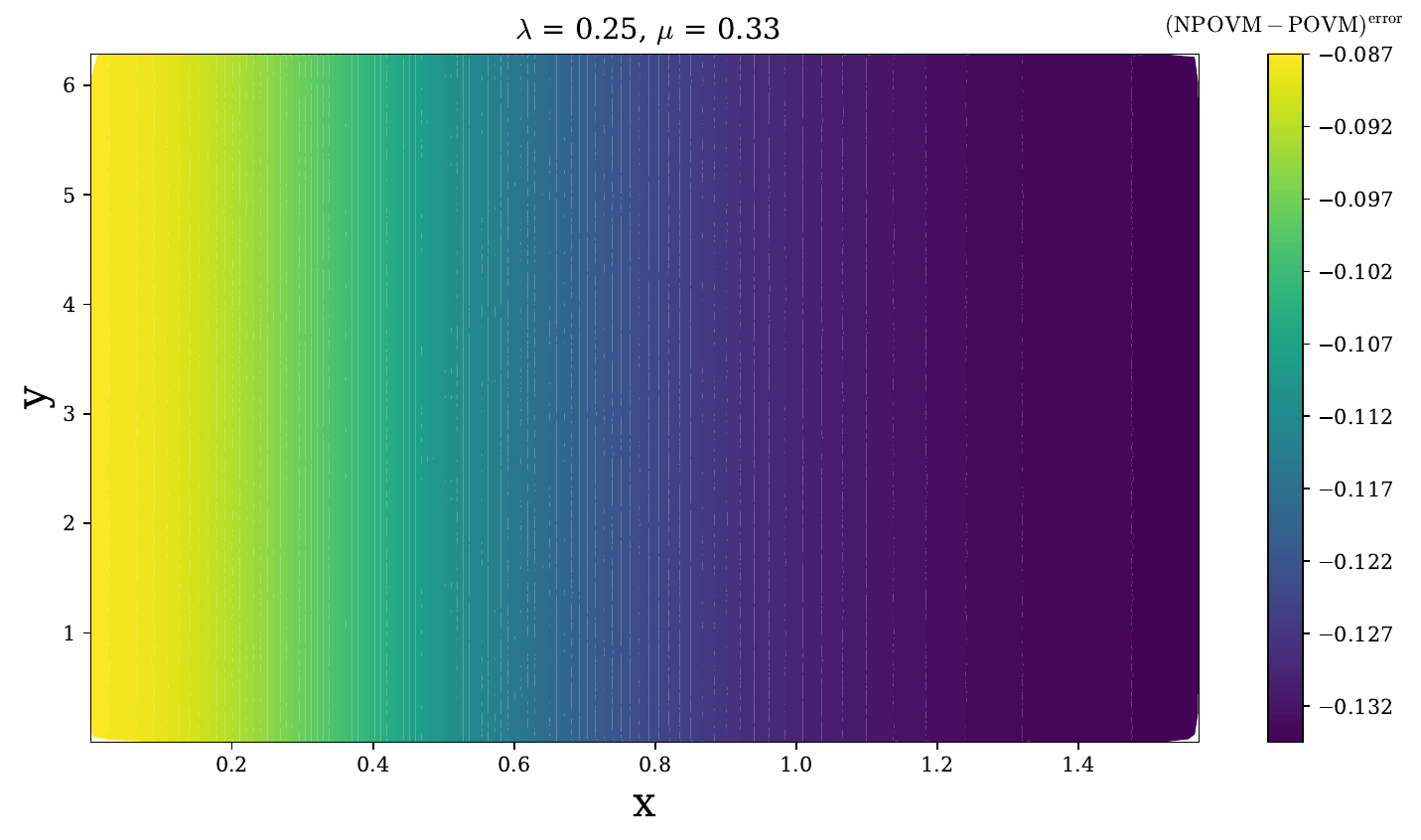}
  \caption{The figure of merit, defined as \( \Delta P_{\text{er}} = P_{\text{error}}^{\text{NPOVM}} - P_{\text{error}}^{\text{POVM}} \), is plotted as a function of the state parameters \( x \) and \( y \) to illustrate the  advantage of NPOVMs over standard POVMs. The analysis pertains to Case IV, wherein the extended joint states are entangled. The plot is generated for fixed values of \( \lambda = 0.25 \), \( \mu = 0.33 \) and $d=0.16$, with all quantities along the axes being dimensionless.
  }
  \label{fig:ncase6}
\end{figure}

In order to evaluate the error performance of NPOVMs under the constrained setting described above, the corresponding optimization problem for the error probability is formulated as
\begin{equation}\label{eq:eqn50}
    \begin{aligned}
        P_{\text{error}}^{\text{NPOVM}} &=  \min \left( \frac{1}{2} - \frac{1}{4} \left\| \rho_{AB} - \sigma_{AB} \right\|_1 \right), \\
        \text{subject to} \quad & \left\{
            \begin{aligned}
               2|\lambda - \mu| &\leq d, \\
               \max \left\{ 2\sqrt{\lambda(1 - \lambda)}, 2\sqrt{\mu(1 - \mu)}) \right\} &\leq E,
            \end{aligned}
        \right.
    \end{aligned}
\end{equation}
where $\rho_{AB}$ and $\sigma_{AB}$ are pure states given in Eqs.~\eqref{eq:eqn40} and~\eqref{eq:eqn41}.
The minimum error probability using POVM can be obtained analytically. The states that we aim to discriminate are $\rho_B$ and $\sigma_B$.
Recall that the difference of the reduced states \(\rho_B - \sigma_B\) is given by
\[
\rho_B - \sigma_B =
\begin{pmatrix}
\lambda - c & -(a - ib) \\
-(a + ib) & c - \lambda
\end{pmatrix},
\]whose 
trace norm evaluates to
$\left\| \rho_B - \sigma_B \right\|_1 = 2 \sqrt{ (\lambda - c)^2 + a^2 + b^2 }$.
The conventional POVM-based error probability is given by
$P_{\text{error}}^{\text{POVM}} = \frac{1}{2} \left[ 1 - \sqrt{ (\lambda - c)^2 + a^2 + b^2 } \right].
$
We calculate the error probability using NPOVM numerically.
In order to compare the POVM-based error with the NPOVM-based one, we fix the parameters to \(\lambda = \frac{1}{3}\) and \(\mu = \frac{1}{4}\) for numerical demonstration and see how  the difference in error probabilities varies as functions of x and y. See Fig.~\ref{fig:case5} for reference. From the figure, we find that in the entire range of values of $x$ and $y$, the figure of merit, i.e. \( \Delta P_{\text{er}} = P_{\text{error}}^{\text{NPOVM}} - P_{\text{error}}^{\text{POVM}} \), gives a negative value, for fixed values of the parameters, $\lambda$ and $\mu$, implying that NPOVM provide an advantage over POVM in discrimainating $\rho_B$ and $\sigma_B$ in this case as well. \hfill $\blacksquare$

As a final scenario, we consider the reduced density matrices on $B$ part to be the same as in theorem~\ref{new}, but instead of the joint states to be entangled, we consider product states of the form given by
\begin{align*}
\rho_{AB} &= \ket{\psi}\bra{\psi}_{A} \otimes \rho_B \;\; \text{and} \\
\sigma_{AB} &= \ket{\phi}\bra{\phi}_{A} \otimes \sigma_B,
\end{align*} 
where  $\ket{\psi}\bra{\psi}_{A}$ and  $\ket{\phi}\bra{\phi}_{A}$ are the same as in case II. Recall that in theorem~\ref{new}, we considered the distance between the auxiliary systems to be $2|\lambda - \mu|\leq d$ . For $\lambda =0.25$ and $\mu=0.33$ we obtained $d=0.16$. For the sake of comparison, here we also fix the constraint at $|\sin(\theta-\phi)|=0.16$ and we then numerically see difference in error probabilities \(P_{\text{error}}^{\text{NPOVM}} - P_{\text{error}}^{\text{POVM}}\) as a function of parameters \(x\) and \(y\). See Fig.~\ref{fig:ncase6} for reference.
From the figure, it is observed that across the entire range of the parameters \( x \) and \( y \), the figure of merit \( \Delta P_{\text{er}} = P_{\text{error}}^{\text{NPOVM}} - P_{\text{error}}^{\text{POVM}} \) remains negative for fixed values of \( \lambda \) and \( \mu \). This consistently negative value indicates that NPOVMs offer a clear advantage over standard POVMs in distinguishing between \( \rho_B \) and \( \sigma_B \) in this setting as well.

\section{Conclusion}\label{sec:Conclusion}
The Helstrom bound represents the fundamental limit on the minimum error probability in binary quantum-state discrimination. 
The Helstrom bound is obtained by minimizing the error probability over the entire set of POVMs. Here we consider a more generalized measurement setting, and minimize the error probability with respect to a set of non-positive operator valued measurements (NPOVMs). To perform an NPOVM on a system, we consider the system to be a part of an extended pure state, and then perform a joint projective measurement on the extended system, followed by tracing out the auxiliary. A crucial feature to note here is that the implementation of NPOVM, in this case, does not necessarily require the extended state to be an entangled state on which the joint projective measurement is to be performed.  We demonstrate that NPOVMs can yield higher success probability of discriminating two states, achieving minimum error rates lower than the conventional Helstrom bound.
So the enhancement in the success probability does not require initial entanglement, and even 
when the joint extended state of Alice and Bob is fully separable, 
non‑POVM strategies can outperform the optimal POVM approach, while discriminating between local states on Bob’s subsystem. 
The state discrimination using NPOVM involves the utilization of resources, i.e. entanglement of the pure-state extensions and partial local distinguishability between the auxiliary states, which constrain the optimization problem of finding the minimum probability of error. The optimization over the set of NPOVMs effectively reduces to a minimization over the auxiliary components of the purified states corresponding to the quantum states under discrimination. 
In conclusion, we investigated several discrimination scenarios, encompassing both the presence and absence of initial entanglement in the joint extended states. Furthermore, we analyzed how initial coherence in the joint states influences the success probability of discriminating between two quantum states using NPOVMs.
\\

\section*{Acknowledgments}

We acknowledge partial support from the Department of Science and Technology,
Government of India, through the QuEST grant with Grant
No. DST/ICPS/QUST/Theme-3/2019/120 via I-HUB QTF of IISER Pune, India. AB acknowledges support from ‘INFOSYS scholarship for senior students’ at Harish Chandra Research Institute, India.
\bibliography{npovm}
\end{document}